# A DATA WAREHOUSE ASSISTANT DESIGN SYSTEM BASED ON CLOVER MODEL

Nouha Arfaoui[1] and Jalel Akaichi[2]

Computer science department, Institut superieur de gestion, 41, Avenue de la liberté, Cité Bouchoucha, Le Bardo 2000, Tunisia

[1] Arfaoui.nouha@yahoo.fr
[2] Akaichi.jalel@isg.rnu.tn

## ABSTRACT

*Nowadays, Data Warehouse (DW) plays a crucial role in the process of decision making. However, their design remains a very delicate and difficult task either for expert or users. The goal of this paper is to propose a new approach based on the clover model, destined to assist users to design a DW. The proposed approach is based on two main steps. The first one aims to guide users in their choice of DW schema model. The second one aims to finalize the chosen model by offering to the designer views related to former successful DW design experiences.*

## 1. Introduction

The data warehousing is becoming increasingly important in terms of strategic decision making through their capacity to integrate heterogeneous data from multiple information sources in a common storage space, for querying and analysis.

In order to fully exploit the DW, it is essential to have a good design to be able to satisfy specified needs and thus give a complete and centralized view of all existing data.

The design is not an easy task because of the necessity to acquire the important knowledge related to the scope and the design techniques used to ensure adequate understanding of the different concepts. Their mastery imposes more effort and time especially with the continuous change and evolution of the domain. This requires to resort to methods such as design methods top-down, buttom-up and middle-out. They greatly facilitate the tasks but despite their utility, they require human intervention each time even if the task has been made by other users.

The idea is to exploit the old manipulations and register them continuously. This provides a rich base of experience. The system, using this base, can guide the user; it can propose solutions and may even predict his future tasks.

The objective of our work is to facilitate the design task for the user and this through the conception and the implementation of a Data Warehouse Assistant Design System based on Clover model (DWADS-Clover). Indeed, our system, in addition to traditional features such as the choice of model diagram of DW (star, snowflake, constellation, etc…), the description of the facts and the sub-attributes underlying, the description of the dimensions and properties, and the establishment of links between fact(s) and dimensions, it tries to exploit the experience of other designers in their borrowing patterns or thereof and propose it to the current user. This is achieved through a component that provides storage and reuse of these experiments. This component is based on the Clover model [1] which uses the paradigm of Case-Based Reasoning (CBR) for excellent grubbing in the data.

DWADS-Clover offers assistance, not related to the use of the tool but rather on the achievement of tasks by providing domain knowledge. It serves to reduce the complexity of tasks while maintaining a certain degree of freedom.

In order to respond appropriately and effectively to its goal, the system must be equipped with a set of characteristics. Indeed, it must be intelligent, i.e. it must be able to ensure automatic registration. It must operate in parallel with the handling of the user. It must register his tasks automatically and continuously. It must also be autonomous, i.e. it intervenes by providing the necessary support without the user's requests through the exploitation of previous experience already registered.

In order to achieve this object, this paper is organized as follows: in the beginning, we talk about works presenting the different design methods as well as the support systems existing in the literature. Subsequently, we present our system based on the meta-model Clover. In the third section, we conduct an implementation accompanied by tests to validate the work. And we end with a conclusion and the prospects of this work.

## 2. State of art

In this part, we start by presenting the different methods related to the DW design, then, we focus on the various support systems exist for make benefit.

In the literature, there is a variety of design methods. The majority of them are based on the model **E**ntity/**R**elationship (**ER**) that reflects, in fact, a natural view of the real world which is modeled through entities and relationships. It has been presented in various works, we cite as examples the following works. [2] considers that this model is more advantageous compared to other models proposed for the databases' design, such as the network model, relational model and the entity set model. It reflects a natural view of the real world which is modeled through entities and relationships. It is used as a basis for a unified view of data. In [3] regarding the construction of the model, it begins by identifying the entities sets and relationships sets that link them, then the ER diagram model is constructed, next the characteristics of all entities and all the relationships that seem to be important towards the user are defined, and finally, the organization of the results obtained at previous levels is done in tables as entity/relationship types and for each set the key is defined. [4] uses this model by associating it with the semantic representation of safety data. Indeed, the ER model provides a graphical representation allowing the description of information, but it is unable to present certain constraints such as security constraints existing in this work. So to overcome this drawback the authors make a combination between this model and a constraint language. According to [5], the use of this model can faithfully present the objects related to the real world and the relationships between them. It has an advantage is that it alleviates the cumbersome computer systems through the construction of small files interconnected by dynamic links. [6] proposes to use the ER method. The latter is based on a set of concept: entity, relationship, role, types and cardinality. These concepts are grouped in a diagram called entity-relationship diagram. The ER model is built based on strong mathematical foundations: the general theory, mathematical relationships, algebra model, etc.

Concerning DW modeling, there are several models but none of them has been considered as a standard [7]. As an example of DW design methods, we quote the **M**ultidimensional **E**ntity-**R**elationship model (**MER**) as presented by [8]. This model is a specialization of the ER. Indeed, it is obtained by adding the concepts related to multi-dimensional paradigm to ER in order to support the modeling of the facts and dimensional levels. We can, then, use a metaphorical cube that corresponds to the analysis subject. The existing cells in this cube contain measures describing the fact, and the axes of the cube represent the different ways for data analysis. [9] present a design model **starER** based both on the star structure which is dominant in the DW, and the ER model which is the most used saw its ability to visualize the elements of the real world as entities linked together by links with a set of attributes. To make the model starER, we begin by gathering user requirements for modeling DW. Another example of work based on the ER model that of [10] who presents a design language called **EV**ent-Entity-Relationship (EVER). The structure of the information at EVER is under the shape of entities and events. An entity is a representation of a mutable phenomenon unique and an event is a unique and immutable representation of an instance of an atomic process of a phenomenon. The event concept is suitable for modeling multidimensional seen that the measured data represents events related to multidimensional databases. In his work, [11] presents a new model of design data in DW known as **G**eneralising **C**onceptual **M**ultidimensional **D**ata **M**odels (**GCMD**). This model is based primarily on the ER model. From the database schema expressed as entity-relationship, he describes the multidimensional structure and this is done by including this dimension with their hierarchical levels. [12] uses the design object-oriented (OO) for the design of multidimensional databases. Indeed, the model **O**bject **O**riented **M**ulti**D**imensional model (**OOMD**) is used. This model is based on two elements namely class size and class of facts which are subsequently presented to the class of cubes as the basic structure for the subsequent analysis of stored data. In [13], the author uses, to model the data, the **profile UML**. In fact, the UML approach visualizes the different ways that a user can use a DW. The purpose is to provide a holistic view that includes all aspects of use of the DW and not only the data modeling. This profile don't present only the multi-dimensional characteristics of database models such as facts, measures, etc, but also it uses advanced features such as degenerate dimensions, etc..[14] presents ad-hoc model based on a traditional databases design and it is composed by the following phases: requirements analysis and specification which serve to present the different information in a ER model database information, conceptual modeling which allows the transmission of semi-formals requirements to a conceptual multidimensional schema visualizing the fact and dimension tables, logical modeling which is used to convert the model obtained above to a logical schema while respecting certain rules and physical modeling which corresponds the implementation of the logical schema while respecting the characteristics of the database. [15] presents a method for developing dimensional models from traditional Entity-relationship models to design Data Warehouses and Data Marts. This method is based on four stages: a first step includes a classification of data model entities in a set of category producing a dimensional model. The second step involves identifying hierarchies presented by a sequence of entity contacted by the one-to-many relationship. The third step is to combine hierarchies and aggregation to form a scale model. The final step includes the evaluation and refinement that requires a combination of fact tables and size tables and resolution of problems arising from many-to-many relations. [16] presents a fact dimensional model. This model starts with the construction of Data Marts. This ensures the success of complex projects. [17] presents a Multidimensional Aggregation Cube (MAC) method. This method provides a construction of a multidimensional schema from the definition of policy needs.

According to the best of our knowledge, there are no models that provide assistance related to DW schema design, but they exist applied in other areas. We can cite as examples, the following works:

In the field of medicine, [18] presents **M**arch **M**edical **A**ssistant (**MMA**). It is an assistant that provides aid to the physician through the provision of information and necessary suggestions. This wizard is based on a combination of three models: the model of user, of situation and of task. In the field of web browsing, [19] presents **Letizia**. It is an agent that assists a user during his navigation through the web or even online information sources. It is based on the history of manipulation in order to anticipate all about user interests. In the same area, [20] presents **Broadway**. It is a counselor that follows the user during his browsing session. It focuses on the current records in order to recommend the following and in order to exploit the ancient browsing. It uses the case based reasoning. [21] presents, in this work, the **A**daptive **P**rogramming **E**nvironment (**APE**). It is an assistant who applies the techniques of machine learning. It is capable of learning the habits of users and subsequently intervenes while suggesting the realization of repetitive tasks. To achieve its mission, APE is composed by three software agents: agent observer, agent learning and agent assistant. These three officers are working in the background. [22] presents **M**odeling **USE**s and **T**asks for **T**racing **E**xperience (**MUSETTE**). This model describes the uses of a system, through the identification of objects and operations. Indeed, all objects observed and manipulated by the user as well as the operations are presented by the use model. The organization of objects and operations over time enables the recovery of trace that can be used to represent a state of the system using objects, or even a transition between two states with successive operations. This trace can be divided into a set of cases called use cases.

## 3. The Clover Model

The Clover model, as presented at [4], was created to enrich the access services to the content of multimedia documents. It is based on three axes namely user, objects and processes (Figure 1).

- The user: it is important to put him into consideration because the Clover model is applied for each user separately. These is why, when he accesses to the application, he must be identified to keep the information related to his manipulation during his session.
- The process: it involves the way to use an element. It represents a point of view related to the manipulation of a computer system.
- The object: as defined in [23], it "is an object that is explicitly manipulated by a computer application".

Indeed, a user logged into the system during his session, handles a set of objects through different processes.

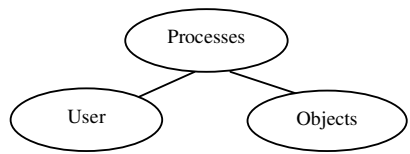

Figure1. The general schema of Clover model **[1]**

### 3.1.    The Clover models

Clover is based on a set of models: use model, observation model and task model. Indeed, the use model focuses on the objects, the rest concentrate on the processes but by different ways.

#### 3.1.1.    The use model

The various objects existed and manipulated by a user are linked together. This connection type is called "contextualization" but this is not the only type of relationship that may exist in computer applications, in fact, we can also find links of composition, membership, etc. The pattern formed by all the related objects is called use model.

According to own application, the objects may be limited. We have thus: the fact tables, dimension tables, links, domains, models, fact attributes, dimension attributes, fact key, and dimension key.

These various objects can be linked into a single generic schema "use model" as presented in the Figure 2. Indeed, we notice that the object model is related to the object domain by the relationship of "contextualization", the same case is applied to the objects table (Fact Table and Dimension Table) which are related to the object model by the same link. This implies that the determination of the model is done in the context of the domain, the determination of tables is done in the context of model, and so on.

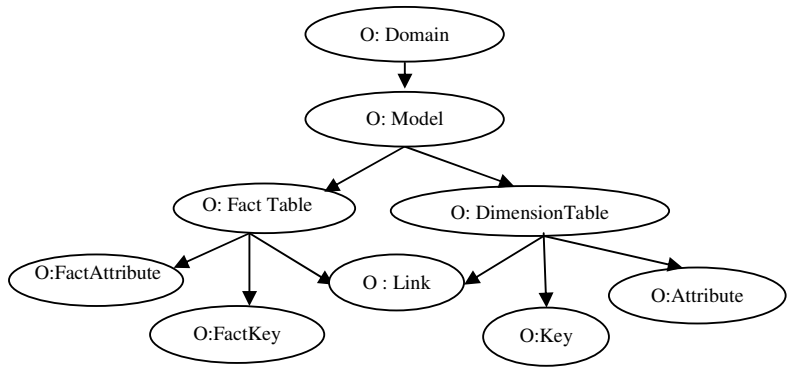

Figure2. Generic use model related to the design of DW

These objects are instancing once the application is used. This instantiation is done thanks to different process such as selecting a domain, selecting a model, creating tables, etc.

Example: The Figure 3 presents a star schema composed by a fact table "Sale" having as primary key "ID-Seller" which refers to the dimension table "Seller" and "ID-Product" which refers to the second dimension table "Product". It has a fact attribute "Sale-Price". The first dimension table "Seller" has a dimension key "ID-Seller" and dimension attribute "Name-Seller". The second dimension table "Product" has a one dimension key "ID-Product" and two dimensions attribute "Name-Product" and "Unit-Price".

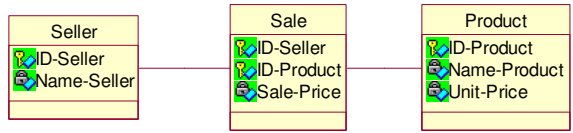

Figure3. Example of a star schema

By applying the generic use model on the example we get the following Figure (Figure 4).

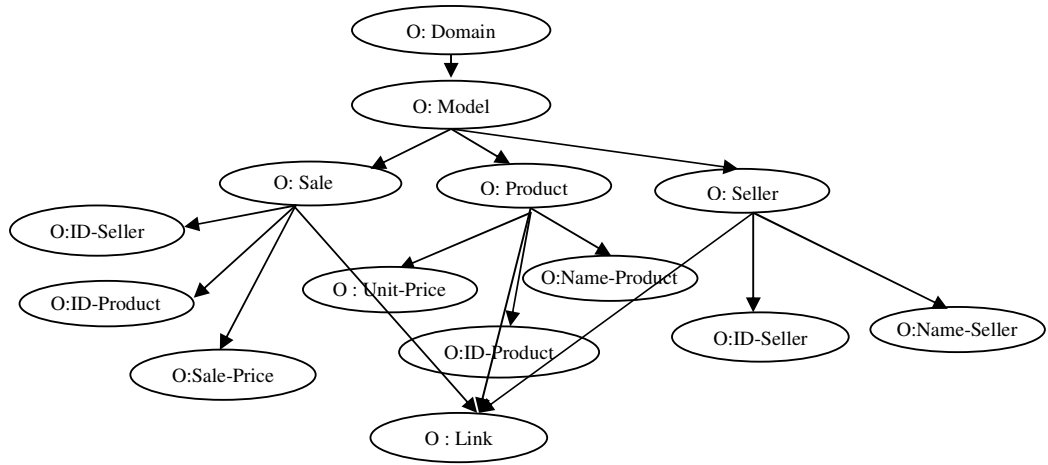

Figure4. Example of the use model

### 3.1.2. Task Model

This model includes all processes applied to objects. Those processes do not imply always an operation or task, but they can be used as processing units. They contain everything ones need to have (objects, operations) to perform a specific task well. They put into consideration all requirements and constraints on an interface.

Related to our application, the creation of a fact table is done through a set of adding key and attributes. Figure 5 presents two different ways allowing an arbitrary decomposition of the task model. Indeed, the process of adding fact attributes is done in the context of the process of adding fact key which is done in the context of creating fact table. Other way, the process of creating table is composed by adding fact key which composed by adding fact attributes (Figure 5 (a)).

Figure 5(b).presents another way for the decomposition. In fact, the two processes adding fact key and adding fact attributes are done in the context of creating fact table, and this last is composed by the two processers adding fact attributes and adding fact key.

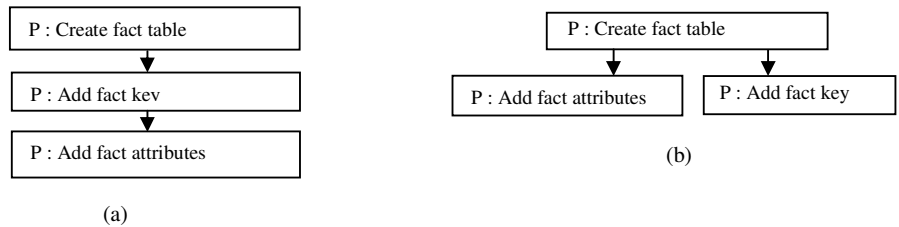

(a)

(b)

Figure5. Task model: example of the creation of a fact table

### 3.1.4. Observation model

This model is constituted by of all processes. It gives a vision on the use and the manipulation of the application. It consists of the entire process handled by a single user during his session.

The Figure 6 presents all processes related to the design of a DW by the user X. In fact, it is a diagram showing the generic observation model related to X.

The design of a scheme is going through a stage set. It begins with the choice of application domain (P: Select the domain) thereafter the choice of a model such as the star, snowflake, etc. (P: Select the model). Once the model is selected, the user heads to create the facts tables (P: Create facts tables), with their keys (P: Add fact key) and attributes (P: Add facts attributes). Subsequently, the dimensions tables are put into consideration (P: Create dimensions tables) with their primary key (P: Add dimension key) and their attributes (P: Add dimensions attributes). Once the tables and their characteristics are defined the user add the links that bind the facts tables and dimensions tables.

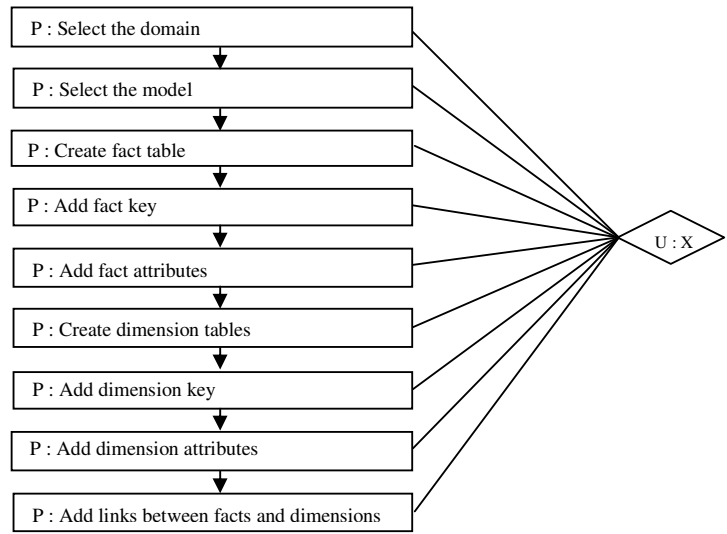

Figure6. Generic observation model related to the design of DW

Example**:** The application of the generic observation model on our example gives us the Figure 7.

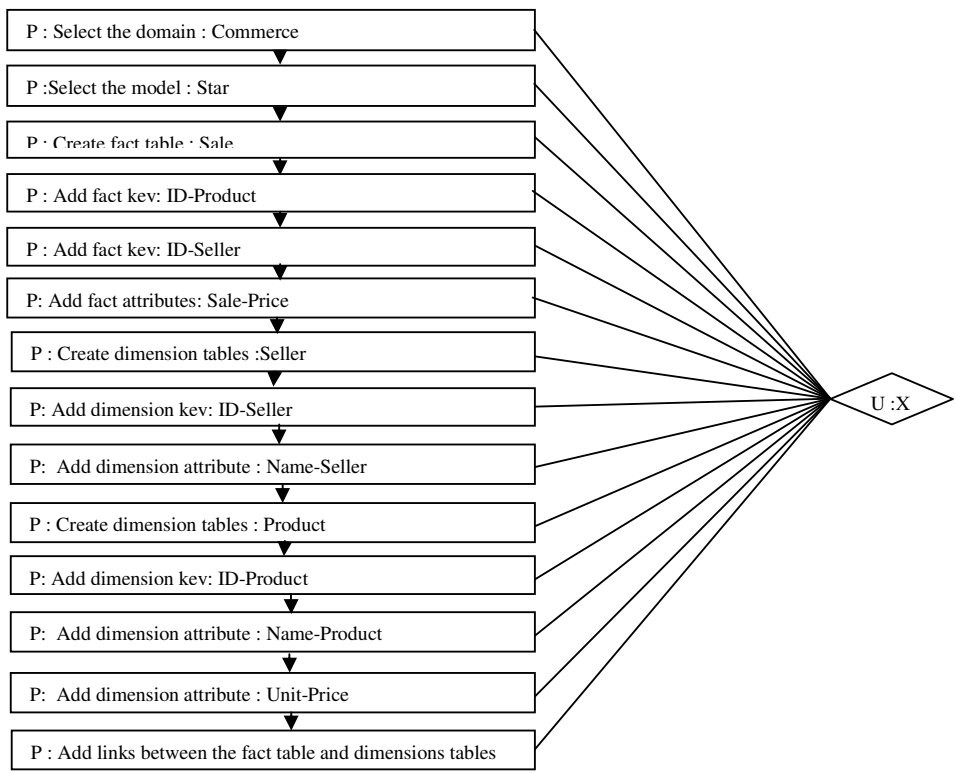

Figure7. Example of the observation model

## 3.2. The graph

Once these models are determined, the system aggregates the use model and the observation model in a single graph to get the gross trace related to the current user. Indeed, a graph consists of a set of nodes and edges. A node may correspond to a user, process or object. It can be abstract or concrete. A concrete node corresponds to the instantiation of an abstract node.

The raw traces presented at the Figure 8 correspond to the manipulation of user X who aims to create a table that has a key and an attribute, and a dimension table with a key and a single attribute.

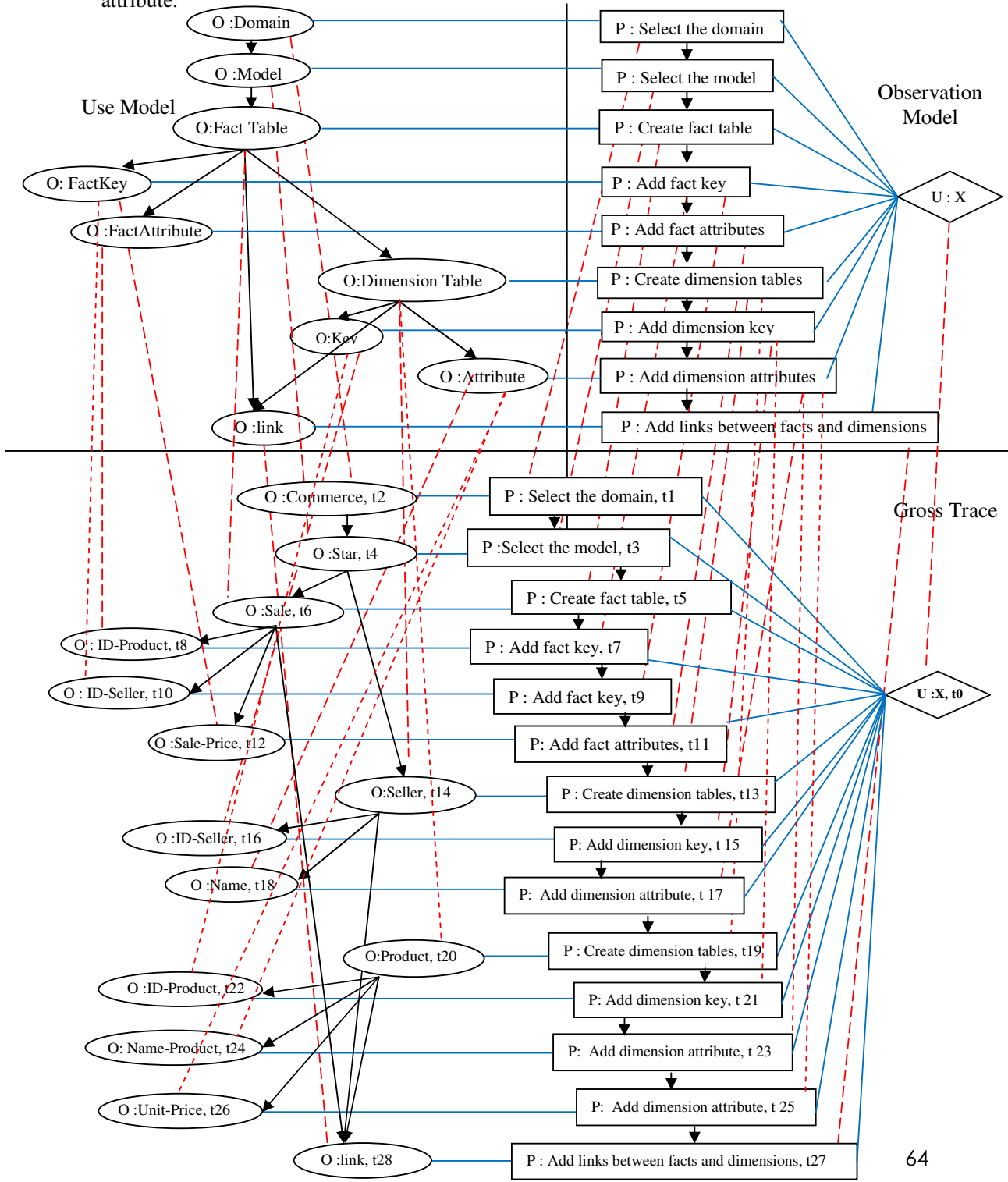

64

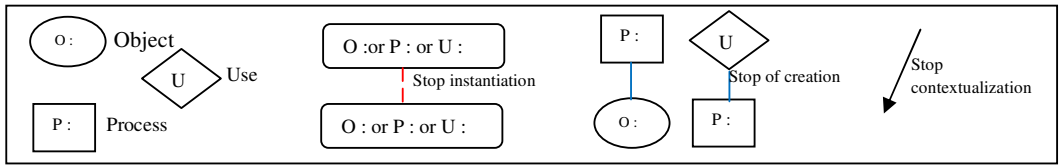

Figure 8. The division of the graph into observation model and use model, and the trace corresponding to the creation of fact table and dimension table

## 4. Trace

The trace can be considered as the basis of our support system. It allows focusing on the ways followed by the user to achieve a well-defined goal. Specifically, it reflects the evolution of both use model and observation model while using the application.

From the Figure 8, we present an example of the construction of a trace. Indeed, we aim to observe the manipulation of abstract objects through abstract processes. This manipulation is related to the abstract user X.

Our user, in Figure 9, performs a single session. It begins with the instantiation of the user node at time t0. This user started by using the process ‖ P: Select the domain ‖ which manipulates the object "O: Commerce", then the process ‖ P: Select the model ‖ with respect to the subject "O: Star" and so on until the obtaining of gross trace corresponding to the instantiation of all abstracts objects and processes. These nodes are arranged in a chronological order.

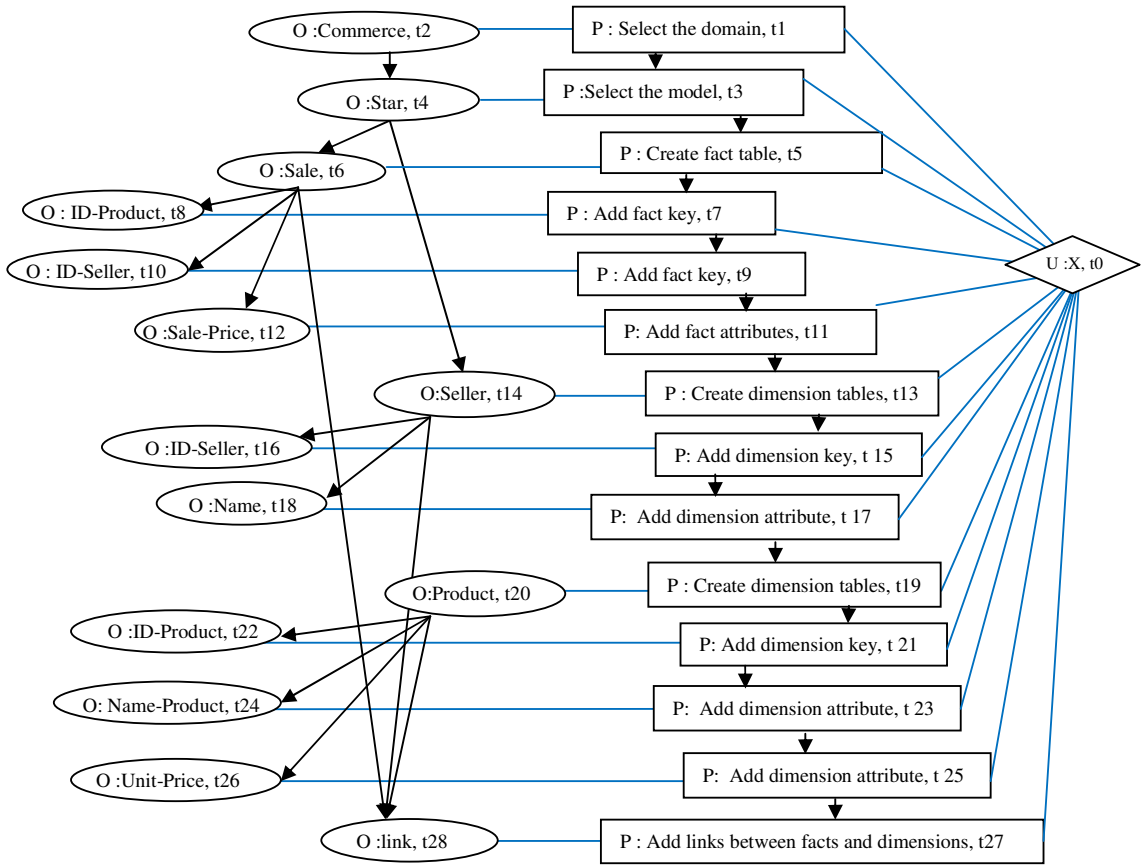

Figure 9. Trace left by the user to create a fact table and dimension table

## 5. Using the experience

Once we get the trace related to the manipulation of the current user, we try to exploit it by extracting the useful information. To realize this purpose, we use the potential graphs. They serve to determine the current level, to predict the next step.

### 5.1. Potential Graphs

The potential graphs are composed by nodes existing at the Clover graph. These nodes are mapped to nodes of the graph overall, while respecting certain elements namely the existing relations, the types of nodes and the similarity between these nodes. They differ from the global graph at the level of the relations. Indeed, other those who exist, there are the temporal relations such as: before, later, during, etc. They have different granularities. Each one allows the determination of a particular object property. At the Clover model, these graphs are used as methods allowing the comparison of the traces related to the current user with those of ancient users. This graph is derived from the use model and contains then the same objects. Figure 10 presents the different granularity existing allowing the consideration of objects.

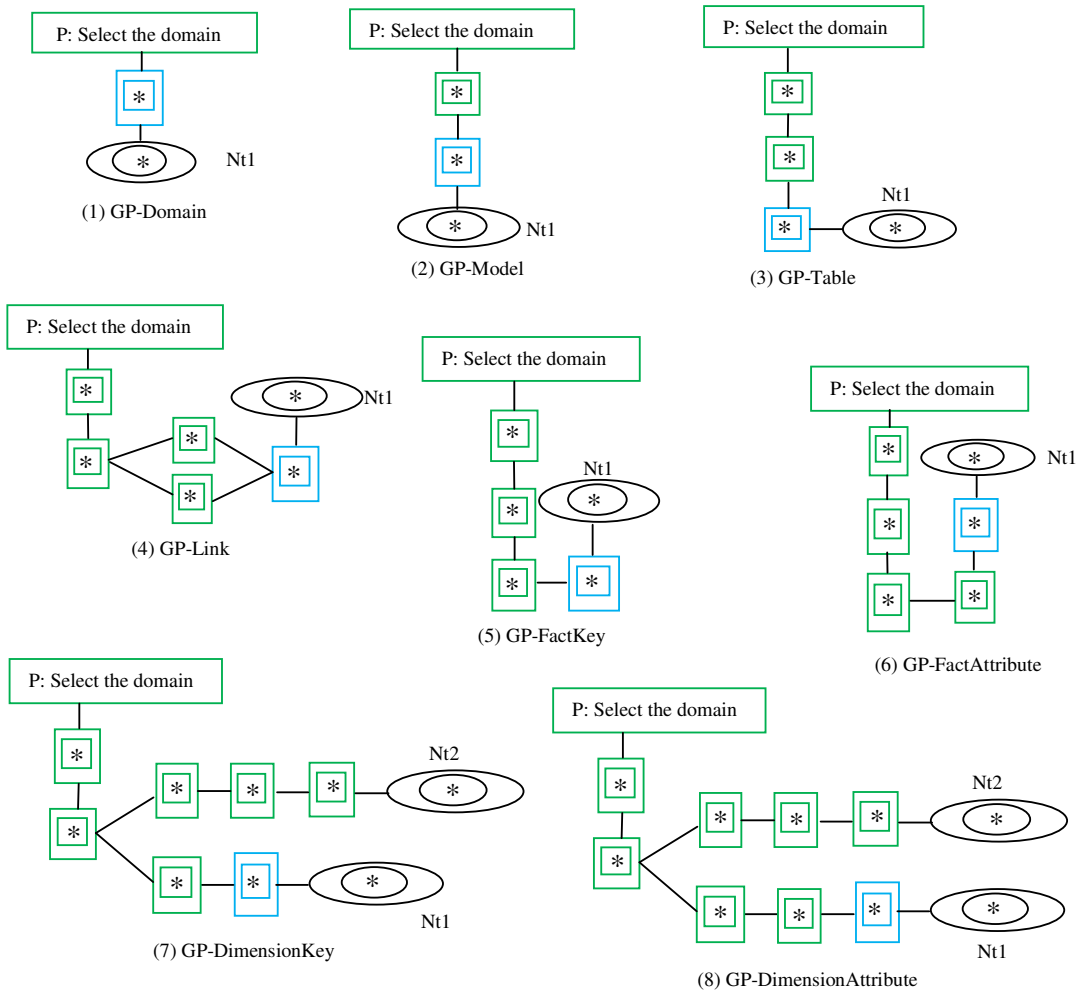

Figure 10. Potential graphs related to the manipulated objects

The Figure 11 shows an example of trace that creates a single fact table with one key and one attribute. This trace is composed by a set of objects and processes. The nodes are extracted from the use and observation models and they are ordered in time. The potential graphs existing on Figure 10 allow calculating this trace.

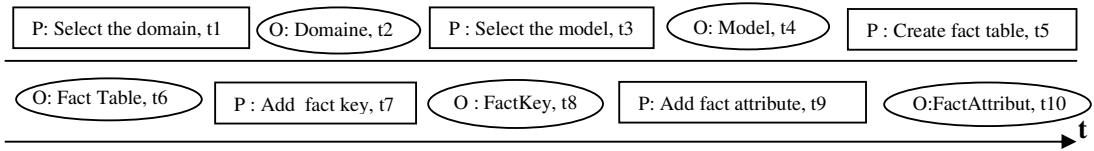

Figure 11. Linear gross trace corresponding to the creation of a fact table with one key and one attribute

Once the system gets the trace corresponding to the manipulation of a user, it uses in two ways. The first serves to update the potential graphs by adding new ones. The second is used to make a set of comparisons in order to determine the current level the user, to know what kind of help he needs to realize his next task.

## 5.2. Clover intervention

The Clover model intervenes by making a set of comparisons which is done automatically and in parallel with the use of the system. This comparison allows detecting the similar cases or even close ones, and in function of the type of similarity, the system proposes the next step.

### 5.2.1. Calculation and comparison

For a new user logged on, the system establishes a set of comparison between his record and those of previous users while using the episodes from the potential graph.
The calculation methods have a thresholds related to the degree of similarity and the minimum number of nodes existing in the target case. Subsequently, we can keep the target having the most cases similar sources for a re-use later.
Through Figure 12, we give an example of episodes calculated on a trace of a user. We present this episode through the observation model. The user, in this case, develops a DW using a star schema. We get then a fact table surrounded by dimension table.
A manner to calculate this episode, it is that presented by Figure 12. It considers the objects belonging to the same level of granularity. By using the potential graphs presented in the Figure 10, we can notice that the instantiation of GP-Area gives the episode E4. GP-Model gives the episode E3, GP-Table and GP-Link give the instantiation of the episode E22 and GP-Key, GP-FactAttribute, GP-DimensionKey, and GP-DimensionAttribute give the instantiation of the episode E1.

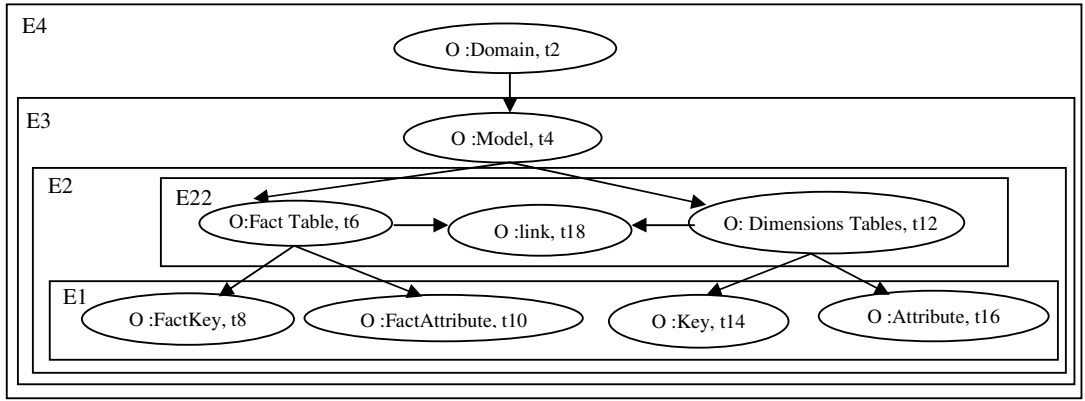

Figure 12. Fitted episodes calculated on a track

Once we have the episodes calculated using the first way or even the second, we pass to the comparison step. It refers to comparable episodes extracted from the traces. Once we have applied the following potential graphs: PG-Table, PG-Link, PG-FactKey, PG-FactAttribute, PG-DimensionKey, PG-DimensionAttribute, we get the traces E2 and E'2 presented on Figure 13.

E2

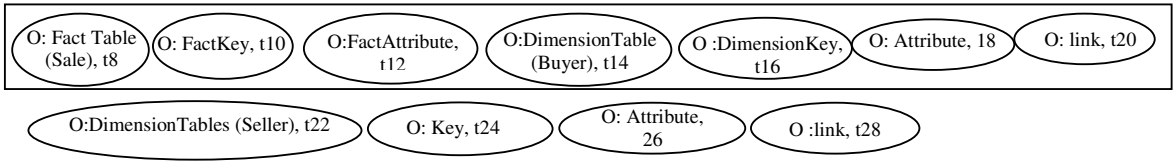

E'2

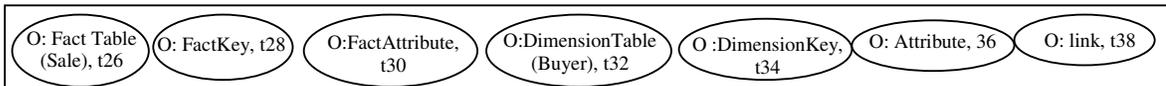

Figure13. Two episodes taken from the traces left by the two sessions

### 5.2.2. Adaptation

For both users, the framed part presents the problem. Related to E2, the rest corresponds to the solution.

Concerning the adaptation, once the system detects this similarity, it may propose to the user Y the continuation of the trace E2.

Generally, the comparison is done using traces with decreasing granularities. In fact, the existing trace at the bottom of the structure has the finest granularity. For our example, ‖ P: Add the links between dimensions and facts tables ‖ is of fine granularity. ‖ P: Add the attributes to dimension table ‖ is less fine granularity and so on until you come to ‖ P: Select the domain ‖.

### 5.2.3. Intervention

The system response depends on the problem part. Indeed, we can distinguish two ways. One way is when this part corresponds to the problem part of another episode. In which case, the system presents the objects of the solution part.

In the case of deference, the system looks in the most similar episodes recorded. In the latter case, it displays all possible cases and, subsequently, gives to the user the freedom to choose.

An episode is considered the most similar, when it contains the last object manipulated by the user, provided that this object is not the last node.

Once the user selects the next object to manipulate the system may intervene again while providing the necessary information regarding the handling of this object. The information is extracted from the task model which contains the constraints of the interface relative to that object.

Information related to the following object: once the system determines the next object, it searches in the task model which allows the extraction of information related to this subject. This serves to give indications to the user regarding how to perform the next task.

## 6. The system architecture

Our system is composed by an interface and an assistant system. The interface presents our system for DW schema design. In fact, it facilitates to the user many tasks such as: selecting a domain, or even adding a new one, handling tables including their attributes and the possible relationship connecting these tables. The assistant system is mainly based on the model of clover as was presented previously. It combines three models (user, observation and task) providing an overview of the actions and their environments. It works in parallel with the user. It can propose, on one hand, the next object which can be manipulated by the user, on the other hand, it records the actions of the current user (Figure 14).

(1) Communication continues with the DW;
(2) Extraction and storage of objects;
(3) Extraction and storage methods;
(4) Extraction and storage tasks;
(5) & (6) Construction of the trace from objects and tasks;
(7) Construction of graph potential graphs (PGs);
(8) Comparison of the trace using the PGs;
(9) Determination of the next object;
(10) Extraction of tasks related to the following object;
(11) Proposal for the next step;
(12) Proposal information about the selected object.

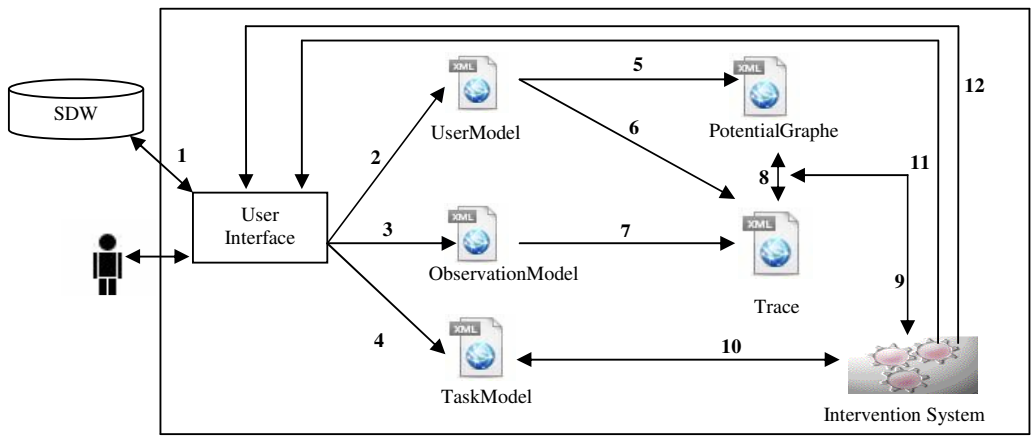

Figure 14. The architecture of DWADS-Clover system

## 7. Conclusion

In this work, we presented the Clover model that allows monitoring and recording traces of users. We have integrated this model into our system of schema design to have a system of aid in the DW schema design.

Indeed, this model is constructed by two parts. The first allows the user tracking through continuous registration of users' traces. This is done through two models: model of observation and use. From these last two we determine the trace corresponding to the instantiation of the objects belonging the two models use and observation.

The trace, on the one hand, allows the determination of the level of the user, on the other hand, it allows the construction of potential graphs which are used as comparison methods for the prediction of possible future paths.

The second part allows the system to intervene by suggesting the next step to take. The procedure is done by consulting ancient manipulations of previous users. At this level, the system exploits the task model to give more information regarding the next object handling.

## ACKNOWLEDGEMENTS

We would like to thank everyone.